\documentclass[Physsubmission, Phys]{SciPost}

\hypersetup{
    colorlinks,
    linkcolor={red!50!black},
    citecolor={blue!50!black},
    urlcolor={blue!80!black}
}

\usepackage[bitstream-charter]{mathdesign}
\DeclareSymbolFont{usualmathcal}{OMS}{cmsy}{m}{n}
\DeclareSymbolFontAlphabet{\mathcal}{usualmathcal}

\begin{document}

\begin{center}{\Large \textbf{
Study of Charmonium(-like) Spectroscopy and Decay at BESIII\\
}}\end{center}

\begin{center}
Weimin Song\textsuperscript{1$\star$}\\
(on behalf of BESIII Collaboration)
\end{center}

\begin{center}
{\bf 1} Jilin University, Changchun, China
\\
* weiminsong@jlu.edu.cn
\end{center}

\begin{center}
\today
\end{center}


\definecolor{palegray}{gray}{0.95}
\begin{center}
\colorbox{palegray}{
  \begin{tabular}{rr}
  \begin{minipage}{0.1\textwidth}
    \includegraphics[width=23mm]{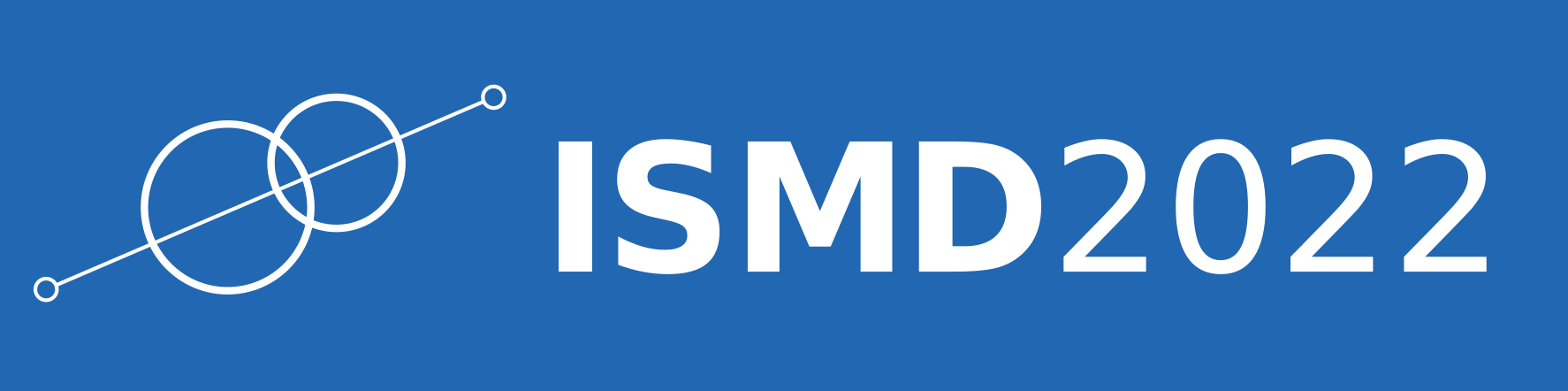}
  \end{minipage}
  &
  \begin{minipage}{0.8\textwidth}
    \begin{center}
    {\it 51st International Symposium on Multiparticle Dynamics (ISMD2022)}\\ 
    {\it Pitlochry, Scottish Highlands, 1-5 August 2022} \\
    \doi{10.21468/SciPostPhysProc.?}\\
    \end{center}
  \end{minipage}
\end{tabular}
}
\end{center}

\section*{Abstract}
{\bf
The recent results on Charmonium and Charmonium-like states at BESIII are reviewed, including the observation of $\rm Z_{cs}(3985)$ state, the study of the new decay modes of $\psi_2(3823)$ state, observation of resonance structure in $\rm e^+e^- \to \pi^+\pi^-\psi_2(3823)$ process, the study of the  $\rm e^+e^- \to K^+K^-J/\psi$ process, cross section measurement of  $\rm e^+e^- \to \omega\pi^0$ and $\omega\eta$ process,  branching fraction measurement of $\psi(3686)\to\bar{\Sigma}^0\Lambda + c.c.$ process. 
Decay channels for $\chi_{cJ} \to \Lambda\bar{\Lambda}$,  $\chi_{cJ} \to \Lambda\bar{\Lambda}\eta$,  $\psi(3686) \to \Lambda\bar{\Lambda}\omega$ and $\eta_c(2S)\to3(\pi^+\pi^-)$ are also discussed.  The property of the spin singlet P wave Charmonium state, $\rm h_c$ is also reported. 
}

\vspace{10pt}
\noindent\rule{\textwidth}{1pt}
\tableofcontents\thispagestyle{fancy}
\noindent\rule{\textwidth}{1pt}
\vspace{10pt}

\section{Introduction}

Just like the spectrum and transitions of the hydrogen atom reveal the physics law between positive charged proton and negative charged electron, which is called Quantum Electrodynamics (QED) nowadays, the spectrum and transitions of Charmonium states will, with no doubt, shed light on the physics law between charm quark and anti-charm quark, which is call Quantum Chromodynamics (QCD). Different from QED, our understanding of QCD is far from complete, especially in the low energy regime where perturbative calculation is not reliable. Charm quark, whose mass scale is just around the limit between the perturbative regime and non-perturbative regime of QCD, is playing a leading role to better understand the QCD. 


The BESIII experiment at the only running electron positron collider in the $\tau$-charm region, Beijing Electron Positron Collider (BEPCII) , has collected the world’s largest Charmonium samples from electron positron symmetric collision. The detector’s performance, such as momentum resolution, particle identification ability, is excellent.
Plenty of results on Charmonium and Charmonium-like states are published by BESIII Collaboration. In this review, the newest results will be presented.

\section{The study of $\rm Z_{cs}(3985)$ state}

After the discovery of the tetra-quark candidate $\rm Z_c(3900)$ state, whose  quark content is believed to be $c\bar{c}q\bar{q}^{\prime}$ ($q^{(\prime)}$ is u or d quark), 
the existence of its strange partner, $\rm Z_{cs}$ with the quark content $c\bar{c}s\bar{q}$, would be expected by assuming the SU(3) flavor symmetry.

\begin{figure}[h]
\centering
\includegraphics[width=0.7\textwidth]{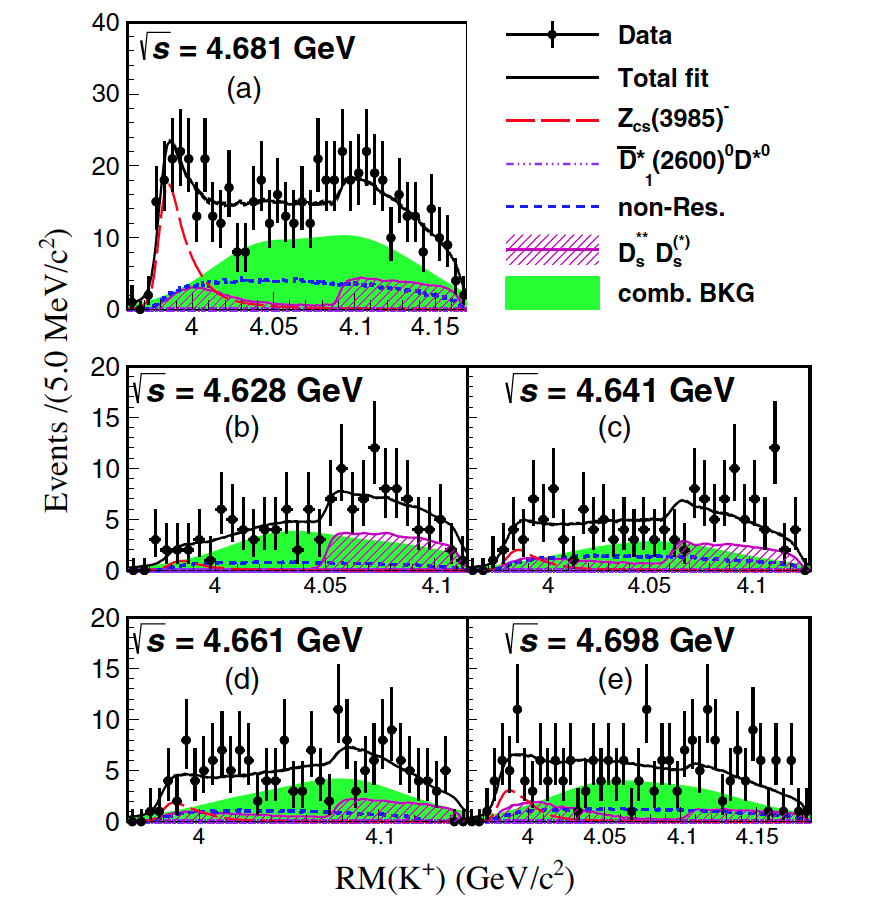}
\caption{Simultaneous unbinned maximum likelihood fit to the $K^{+}$ recoil-mass spectra in data at 4.628, 4.641, 4.661, 4.681 and 4.698 GeV.}
\label{ref3985}
\end{figure}

 When studying the $e^+e^- \rightarrow K^+(D_s^-D^{*0}+D_s^{*-}D^0)$ process with data corresponding to integrated luminosity of  3.7~$\rm fb^{-1}$ at five center-of-mass energies ranging from 4.628 to 4.698~GeV, an excess of events is observed in the $K^{+}$ recoil-mass spectrum, as shown in Fig.~\ref{ref3985}~\cite{BESIII:2020qkh}. The significance of the structure is 5.3$\sigma$, and its mass and width are measure to be $(3982.5^{+1.8}_{-2.6}\pm2.1)$~MeV$\rm /c^2$ and $(12.8^{+5.3}_{-4.4}\pm3.0)$~MeV.

With the same dataset, the isospin partner is found with a strong evidence (4.6$\sigma$) in the recoil mass spectra of $K^0_S$ in $e^+e^- \rightarrow K^0_SD_s^+D^{*-}$ and  $e^+e^- \rightarrow K^0_SD_s^{*+}D^{-}$ process~\cite{BESIII:2022qzr}. LHCb collaboration observed a state in the $J/\psi K^+$ invariant mass spectrum~\cite{LHCb:2021uow}, namely the $\rm Z_{cs}(4000)$ state, however the width is very different from the one measured at BESIII.

\section{The study of $\rm \psi_2(3823)$ state}

The $1^3D_2$ Charmonium state, $\rm \psi_2(3823)$, was observed by BESIII Collaboration in $\gamma \chi_{c1}$ final state, however the experimental information about it is still poor. With data sample corresponding to integrated luminosity of 19 $\rm fb^{-1}$  at center-of-mass energies between 4.1 and 4.7~$\rm GeV$, the decay processes of $\psi_2(3823) \to \gamma \chi_{c0,1,2}$, $\pi\pi J/\psi$, $\eta J/\psi$ and $\pi^0 J/\psi$ are searched for~\cite{BESIII:2021qmo}.  The $\psi_2(3823) \to \gamma \chi_{c1}$ is re-discovered with higher significance, and  an evidence for $\psi_2(3823) \to \gamma \chi_{c2}$ process is found for the first time with a significance of $3.2 \sigma$. The ratio between these two decay modes is consistent with the predictions. No significant signals are observed for the other decay modes.  No significant $\rm e^-e^+ \to \pi^+ \pi^- \psi_3(3842)$ signals are seen in any of the channels we studied. The process $\rm e^-e^+ \to \pi^0 \pi^0 \psi_3(3823)$ with $\psi_2(3823) \to \gamma \chi_{c1}$   is also searched for, and evidence for the process is found with a significance of 4.3$\sigma$.

\begin{figure}[h]
\centering
\includegraphics[width=0.7\textwidth]{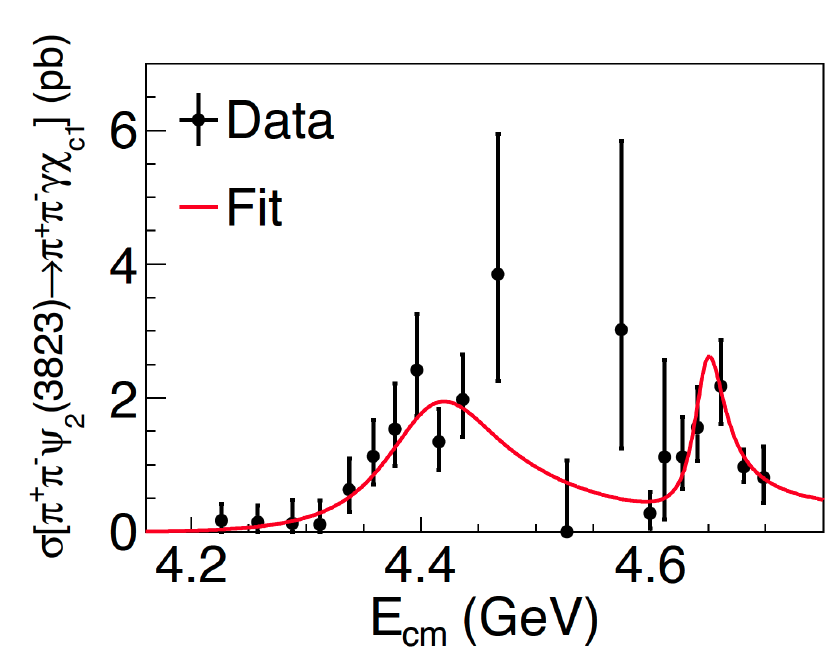}
\caption{Result of the fit to the $\sqrt{s}$ dependent cross section. Dots with error bars are data, and red solid curve shows the fit with two coherent resonances.}
\label{ref3823}
\end{figure}

As shown in Fig.~\ref{ref3823}, the product of the $\rm e^-e^+ \to \pi^+ \pi^- \psi_3(3823)$ cross section and the branching fraction of $\psi_2(3823) \to \gamma \chi_{c1}$  is measured as a function of the center-of-mass energy~\cite{BESIII:2022yga}. This is the first observation of vector Y-states decaying to D-wave Charmonium state, which provides new insights about the Y -states wave functions. Within current uncertainties, the resonance parameters of the two structures are similar to the Y(4360) and Y(4660) state. With the same data, we performed the most precise mass measurement and provided most stringent constraint on the width of $\psi_2(3823)$ state by taking $\rm e^-e^+ \to \pi^+ \pi^- \psi(3686)$ process as reference channel. 

\section{The study of vector Charmonium-like states}

A few vector Charmonium-like states, such as Y(4260), Y(4360) and Y(4660) were observed mainly in Charmonium plus light hadrons final state, and in order to understand the nature of these states, their pure light hadron final states are studied also, however no significant signal is observed yet. 

\begin{figure}[h]
\centering
\includegraphics[width=0.7\textwidth]{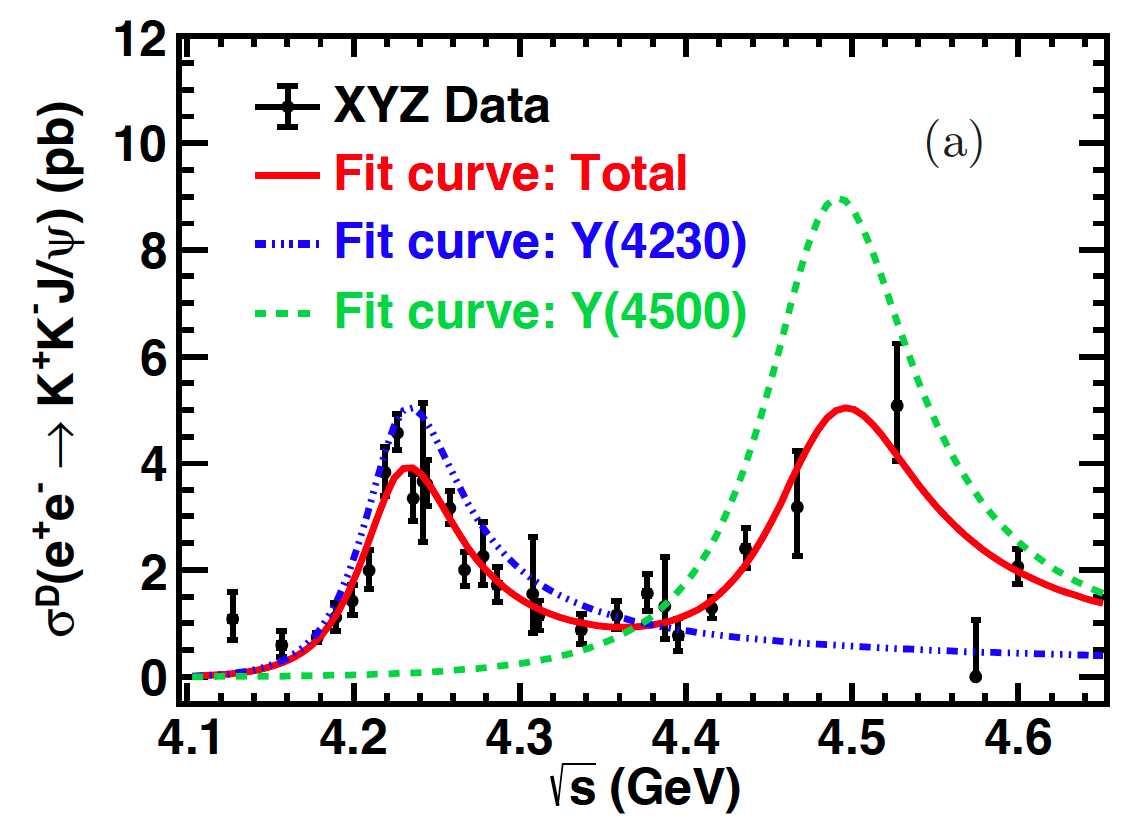}
\caption{Result of the fit to the $\sqrt{s}$ dependent cross section for $e^+e^- \rightarrow K^+K^-J/\psi$ process. }
\label{refkkjpsi}
\end{figure}

As shown in Fig.~\ref{refkkjpsi}, the Born cross section of $e^+e^- \rightarrow K^+K^-J/\psi$ process is measured with a new partial reconstruction method and larger data samples compared with the previous study~\cite{BESIII:2022joj}.  Two resonances are observed with high significances, and the first one is consistent with Y(4230), the second one is a new structure which is named as Y(4500). The nature of Y(4500) is still not clear. For the first time, the state Y(4230) has been observed in the $K^+K^-J/\psi$ mode with the significance larger than 5$\sigma$. As multiple solutions exit, it is not easy to make any conclusion about the nature of Y(4230) based on the information about its decay partial widths.

Among the light hadron final state, the one with proton and anti-proton in the final state is of more interest when searching for the exotic states, because this information is valuable for estimating the production rates of these exotic states on PANDA experiment. Recently, BESIII measured the cross section of $e^+e^- \to p\bar{p}\eta$ and $e^+e^- \to p\bar{p}\omega$, and no clear structure is observed on the cross section line shape~\cite{BESIII:2021vkt}. It is same for $e^+e^- \rightarrow \omega\pi^{0}$ and $\omega\eta$ process, no structure is found on the cross section line shape~\cite{BESIII:2022zxr}.  

\section{The study traditional Charmonium states}

\begin{figure}[h]
\centering
\includegraphics[width=0.45\textwidth]{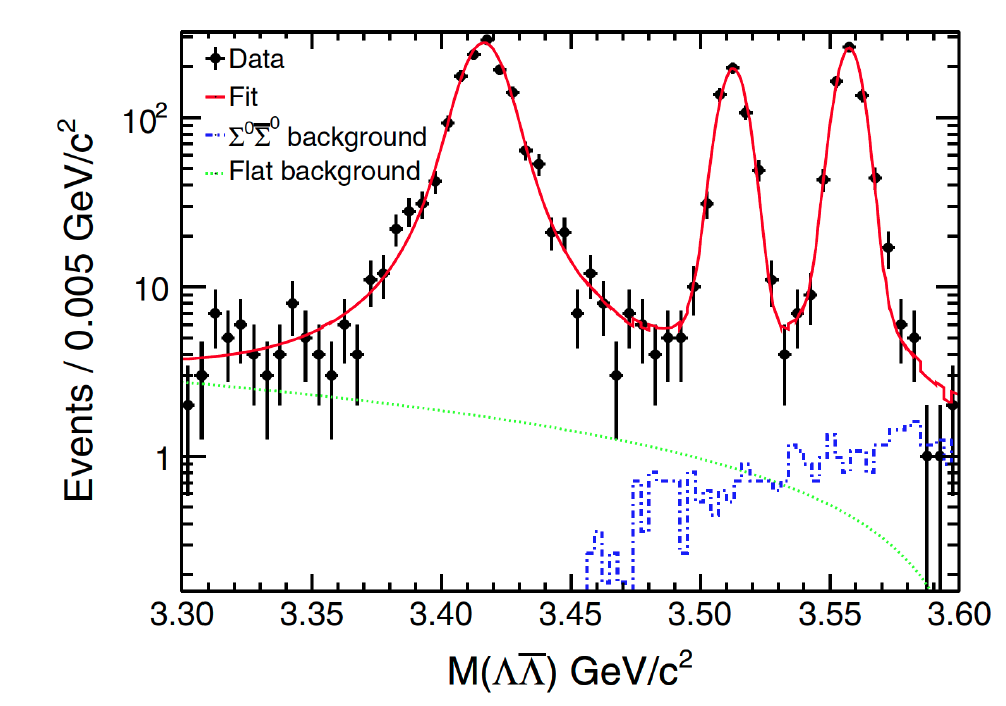}
\includegraphics[width=0.45\textwidth]{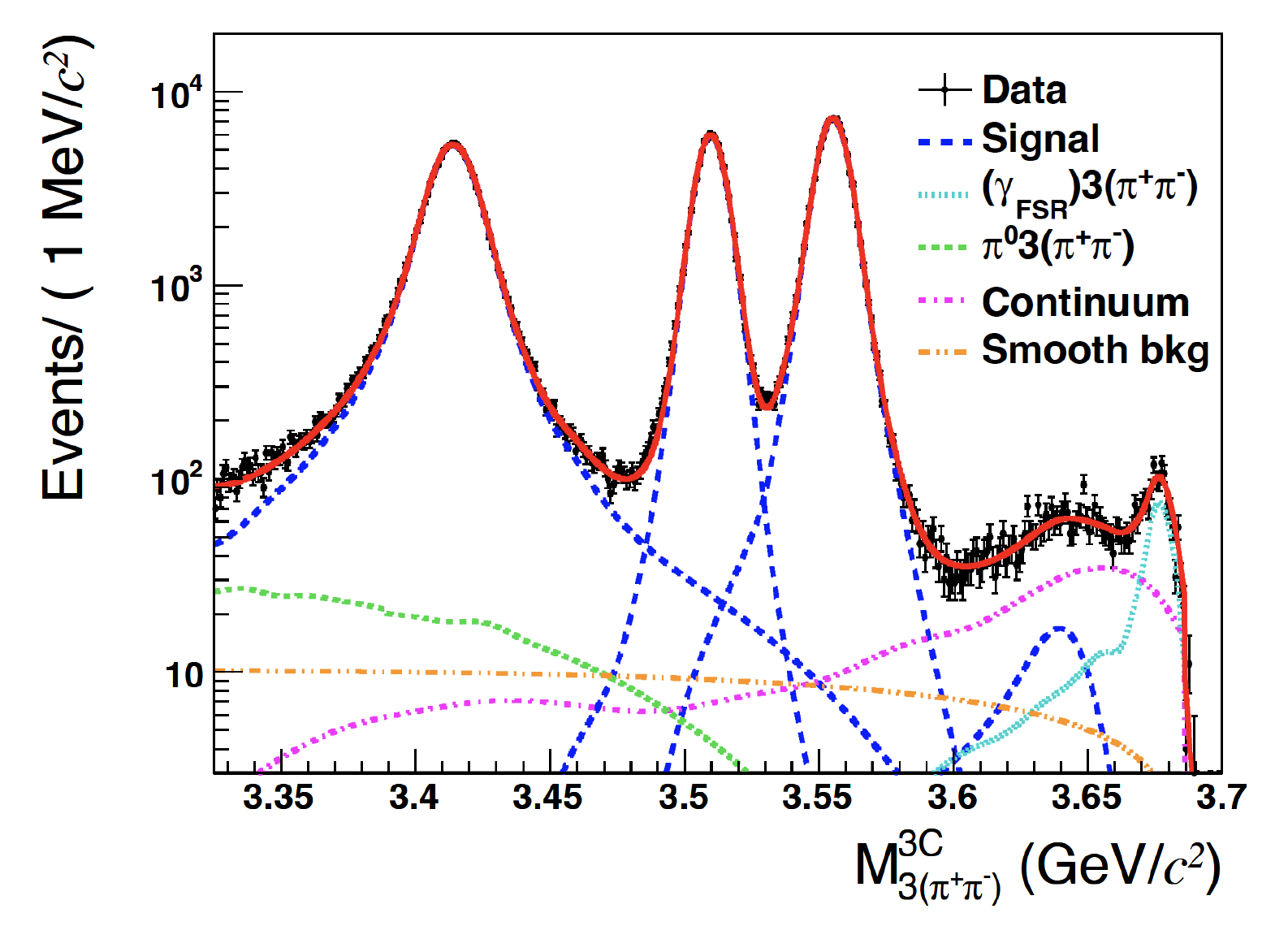}
\caption{The fit to the invariant mass of $\Lambda \bar{\Lambda}$ in $\chi_{cJ (J = 0,1,2)} \rightarrow \Lambda \bar{\Lambda}$ (left); The fit to the invariant mass of $3(\pi^+\pi^-)$ in $\chi_{cJ} \to 3(\pi^+\pi^-)$ (right).}
\label{refchicj}
\end{figure}

The Charmonium states below the open charm threshold include $J/\psi$, $\psi(3686)$, $\eta_c$, $\eta_c(2S)$, $\chi_c(0,1,2)$ and $h_c$ state, and all of these could be studied with the electron positron collision data taken at BESIII. In principe, only the ones with quantum number $J^{PC}=1^{--}$ could be produced with large rate as they have the same quantum number as the photon. The others could be produced by radiative transitions or hadronic transitions.  As shown in Fig.~\ref{refchicj}, the process of $\chi_{cJ (J = 0,1,2)} \rightarrow \Lambda \bar{\Lambda}$  and $\chi_{cJ} \to 3(\pi^+\pi^-)$ could be studied at large samples~\cite{BESIII:2021mus}\cite{BESIII:2022hcv}.  A handful of other decay channels, such as $\chi_{cJ}\to \Lambda\bar \Lambda \eta$~\cite{BESIII:2022oiz}, $\psi(3686)\rightarrow\Lambda\bar{\Lambda}\omega$~\cite{BESIII:2022fhe} process are also studied.  The mass, width, and branching fractions for $h_c$ are studied precisely via the decay $\psi(2S) \rightarrow \pi^0 h_c$~\cite{BESIII:2022esq}.

\section{Conclusion}

Using world's largest electron-positron annihilation data in the charm region, the Charmonium(like) sates could be studied with high precision. 
With these results, we hope to understand the QCD better, especially in the low energy region. 

\section*{Acknowledgements}
The author would like to thank all the colleagues in the BESIII Collaboration for their great effort to make the mentioned results available, and of course thank the organiser of ISMD2022 conference to make the extraordinary event happen. 

\paragraph{Funding information}
This work is supported by Program of Science and Technology Development Plan of Jilin Province of China under Contract No. 20210508047RQ.




\nolinenumbers

\end{document}